\documentclass[prc,twocolumn]{revtex4}
\usepackage{amsmath,amssymb,epsfig,bm,mathrsfs,feynmp,slashed,color,isotope}
\usepackage[colorlinks=true,linkcolor=blue,citecolor=blue,urlcolor=blue]{hyperref}
\usepackage{nicefrac}
\usepackage{graphicx}
\usepackage{exscale} 
\usepackage{gensymb}
\usepackage{slashed}
\usepackage{isotope}
\usepackage{color}
\newcommand{\be}{\begin{equation}}
\newcommand{\ee}{\end{equation}}
\newcommand{\bea}{\begin{eqnarray}}
\newcommand{\eea}{\end{eqnarray}}

\newcommand{\vecS}{{\bm S}}

\newcommand{\vecB}{{\bm B}}
\newcommand{\vecL}{{\bm L}}

\newcommand{\vecl}{{\bm l}}
\newcommand{\vecsigma}{{\bm \sigma}}

\newcommand{\sky}{{\sc \footnotesize  SKY3D }}

\definecolor{red}{rgb}{0.8,0,0}
\definecolor{orange}{rgb}{0.8,0.2,0.0}
\definecolor{blue}{rgb}{0.3,0.0,0.8}

\begin{document}

\title{Carbon-oxygen-neon mass nuclei in superstrong magnetic fields}

\author{Martin Stein}
\thanks{\tt mstein@th.physik.uni-frankfurt.de}
\affiliation{Institute for Theoretical Physics,
  J.~W.~Goethe-University, D-60438  Frankfurt am Main, Germany}

\author{Joachim Maruhn}
\thanks{\tt maruhn@th.physik.uni-frankfurt.de}
\affiliation{Institute for Theoretical Physics,
  J.~W.~Goethe-University, D-60438  Frankfurt am Main, Germany}

\author{P.-G. Reinhard}
\thanks{\tt paul-gerhard.reinhard@physik.uni-erlangen.de}
\affiliation{
Institut f\"ur Theoretische Physik, Universit\"at Erlangen, D-91054 Erlangen, Germany
}

\author{Armen Sedrakian}
\thanks{\tt sedrakian@th.physik.uni-frankfurt.de}
\affiliation{Institute for Theoretical Physics,
  J.~W.~Goethe-University, D-60438  Frankfurt am Main, Germany}

\date{\today}

\begin{abstract}
  The properties of $\isotope[12]{C}$, $\isotope[16]{O}$, and
  $\isotope[20]{Ne}$ nuclei in strong magnetic fields
  $B\simeq 10^{17}\,$G are studied in the context of strongly
  magnetized neutron stars and white dwarfs. The \sky code is
  extended to incorporate the interaction of nucleons with the
  magnetic field and is utilized to solve the time-independent
  Hartree-Fock equations with a Skyrme interaction on a Cartesian three-dimensional
  grid. The numerical solutions demonstrate a number of phenomena,
  which include a splitting of the energy levels of spin-up and -down
  nucleons, spontaneous rearrangment of energy levels in
  $\isotope[16]{O}$ at a critical field, which leads to jump-like
  increases of magnetization and proton current in this nucleus,
  and   evolution of the intrinsically deformed $\isotope[20]{Ne}$ nucleus
  towards a more spherical shape under increasing field strength. Many
  of the numerical features can be understood within a simple analytical
  model based on the occupation by the nucleons of the lowest states
  of the harmonic oscillator in a magnetic field.
 \end{abstract}

\maketitle

\section{Introduction}
\label{sec:intro}

The studies of nuclei and bulk nuclear matter in strong magnetic
fields are motivated by the astrophysics of strongly magnetized
neutron stars (magnetars) and white dwarfs. The
surface fields of magnetars have been inferred to be from observations in
the range of $10^{15}$~G. The interior fields of magnetars can be
several orders of magnitude larger than their surface
fields~\cite{2015RPPh...78k6901T}.

The electromagnetic energy of the interaction of baryons with the
$B$ field becomes of the order of the nuclear scale $\sim$ MeV for fields
$10^{16}-10^{17}$~G and can arise from current-field (for electrically
charged particles) and spin-field (for charge neutral particle)
interaction. Such interaction can affect the properties of nuclei,
including their shell structure, binding energies, and rms radii.
 This, in turn, can affect the structure and composition of the
interiors of neutron stars and white dwarfs where nuclei are predicted
to exist, as well as the transport and weak interaction (neutrino
emission and absorption) processes due to the changes in charged
particle dynamics and transition matrix elements.

The equation of state and composition of inhomogeneous nuclear matter
featuring nuclei in strong magnetic fields have been studied using
various methods including modification of the Thomas-Fermi model 
~\cite{1989ApJ...342..958F}, liquid drop
model~\cite{1991ApJ...383..745L}, nuclear shell Nilsson
model~\cite{2000PhRvL..84.1086K,2001ApJ...546.1137K}, relativistic
density functional
theory~\cite{2011PhRvC..84d5806P,2015PhRvC..92c5802B}, and
non-relativistic Skyrme functional theory~\cite{2012PhRvC..86e5804C}. It has
been shown that bulk properties of nuclei and shell structure as well
as their shape can be significantly affected if the magnetic field is of
the order of $10^{17}$~G and larger. These studies were carried out in
the context of neutron star crusts and have concentrated on heavy
nuclei beyond (and including) $^{56}$Fe. 

 In this work we consider the properties of carbon-oxygen-neon mass
 nuclei in strong magnetic fields. Our motivation for doing so is
 threefold. Firstly, some isolated neutron stars are known to have
 atmospheres composed of $\isotope[12]{C}$, as is well established in
 the case of the compact object in Cas
 A~\cite{2013ApJ...779..186P}. Carbon plays also an important role in
 the physics of accreting neutron stars, for example, in the
 superbursts which are associated by unstable ignition of carbon at
 depth characterized by the density $\rho\simeq 10^9$ g\,
 cm$^{-3}$~\cite{2014ApJ...791..106S}.  Apart from $\isotope[12]{C}$,
 there are also substantial fractions of oxygen and neon mass nuclei
 produced in the crusts of accreting neutron stars through nuclear
 reaction networks~(Ref.~\cite{2014ApJ...791..106S} Table I and
 Figs. 4 and 5). The composition of the surfaces of magnetars and
 their properties under accretion are not known. However, one can
 anticipate the role played by these nuclei in the physics of
 magnetars, by extrapolating from the physics of low-field neutron
 stars. Secondly, white dwarfs models composed of $\isotope[12]{C}$,
 $\isotope[16]{O}$, or $\isotope[20]{Ne}$ are standard in the physics
 of these compact objects in the non-magnetized regime. The superluminous
 type-I supernovae were suggested recently to originate from
 supermassive strongly magnetized white
 dwarfs~\cite{2013IJMPD..2242004D} with magnetic fields in the range
 $10^{17}$~G.  If such objects exist (for a discussion 
see~\cite{2014PhRvD..90d3002C}) they
 would provide the environment where light nuclei would be subjected to
 intense $B$ fields.  Thirdly, besides the astrophysical motivation,
 there is a technical aspect to our study, as it is a first attempt to
 include magnetic fields in the widely used \sky
 code~\cite{2014CoPhC.185.2195M}. Therefore, another motivation of our
 study is to provide a benchmark for future studies that will include
 magnetic fields on simple enough systems that are easily tractable,
 and $\isotope[12]{C}$, $\isotope[16]{O}$, or $\isotope[20]{Ne}$ nuclei
 are optimal in this respect. As we show below, the splitting of the
 levels induced by the $B$ field in these nuclei can be easily
 understood on the basis of a harmonic oscillator model; because of
 the small number of levels the comparision between the numerical and
 analytical results becomes possible. (Indeed, in heavy nuclei the
 number of levels can become very large, which would make such
 comparison impractical).

The density-functional-based Hartree-Fock  (HF) theory provides an
accurate and flexible framework to study a variety of low-energy
nuclear phenomena. The public domain  \sky
code~\cite{2014CoPhC.185.2195M}, solves the HF equations on a 3D grid
(i.e. without any assumptions about the underlying symmetries of the
nuclei) and  is based on the Skyrme density functional.  It has
already been utilized to study a broad range of problems ranging from
low-energy heavy-ion collisions to nuclear structure to exotic shapes
in crusts of neutron stars (for references see
Ref.~\cite{2014CoPhC.185.2195M}).

In this work we report on the first implementation of a strong magnetic
field in the \sky code via extension of the Hamiltonian (and the
associated density functional) to include all relevant terms
reflecting the interaction of the magnetic field with nucleons. We
concentrate on the static solutions, which requires the solution of
time-independent Hartree-Fock equations.  As an initial application we
report convergent studies of the carbon-oxygen-neon mass range nuclei
in strong fields.

This paper is organized as follows. In Sec.~\ref{sec:theory} we
describe briefly the underlying theory and the modifications to the
numerical code needed to include $B$ fields. We present our results in
Sec.~\ref{sec:results} where we first set up a simple analytical model
which is then compared with numerical results. A number of observables
such as energy levels, spin- and current-densities, as well as
deformations are discussed. Our conclusions are summarized in
Sec.~\ref{sec:conclusions}.

\section{Theory and numerical code }
\label{sec:theory}

\subsection{Theory}

We consider a nucleus in a magnetic field which is described by the
Hamilton operator $ \hat h_q = \hat h^{(0)}_q +\hat h^{(B)}_q$, where
$q\in\{p,n\}$ specifies the isospin of a nucleon and $\hat h^{(0)}_q $
is the Hamiltonian in the absence of a $B$ field, which is given by Eq. (8)
of Ref.~\cite{2014CoPhC.185.2195M}. It is constructed in terms of
densities and currents of nucleons.  The term $\hat h^{(B)}_q$ which
is due to the $B$ field is given by
\begin{eqnarray}
\label{sphmod}
\hat h^{(B)}_q  &=& -\left(\vecl\delta_{q,p} 
+ g_q\frac{\vecsigma}2\right) \cdot \tilde\vecB_q,
\end{eqnarray}
where $\tilde{\vecB_q}\equiv ({e\hbar}/{2\,m_qc})\vecB\,,$ 
$\vecsigma$ is the spin Pauli matrix, and $\vecl$ is the
(dimensionless) orbital angular momentum. These are related to the
spin $\vecS$ and the orbital angular momentum $\vecL$ via
$\vecS=\hbar\vecsigma/2$ and $\vecL=\hbar\vecl$, where $g_n=-3.826$
and $g_p=5.585$ are the $g$ factors of neutrons and protons.  The
Kronecker-delta takes into account the fact that neutrons do not
couple with their orbital angular momenta.

Starting from the Hamiltonian $\hat h_q $ one can construct the
density functional (DF) in the basis Hartree-Fock wave functions. We
use the DF which contains the kinetic energy, the nuclear Skyrme
interaction, the Coulomb energy, and the correlation correction to the
mean-field DF. In principle the pairing correction can be added to the
DF, but we ignore it here because of its overall smallness and because
the $S$-wave pairing would be quenched by the magnetic
field~\cite{2016PhRvC..93a5802S}.  The correlation correction in the
DF includes those corrections which reflect the beyond-mean-field
contributions.

\subsection{Numerical code and procedure}

We have used the code \sky to find iteratively the solution of the static
HF equation in strong magnetic fields.  The equations are solved using
successive approximations to the wave functions on a three-dimensional
Cartesian grid, which has 32 points along each direction and a
distance between the points of 1.0 fm. (We varied the parameters of the
mesh within reasonable values and made sure that the physical results
are unchanged).  To accelerate the iteration a damping of the kinetic
term is performed.  The following ranges of the two numerical
parameters of the damped gradient step have been used - the step size
$\delta = 0.4$ and the damping regulator $E_0 = 100$ MeV.
After performing a wave-function iteration steps, the densities of the
nucleons are updated and new mean fields are computed.  This provides
the starting point for the next iteration. The iterations are
continued until  convergence to the desired accuracy is achieved. As
a convergence criterion the average energy variance or the fluctuation
of single-particle states is used. To initialize the problem we use
harmonic oscillator states and include an unoccupied nucleon state.  The
radii of the harmonic oscillator states are $r_{\rm h. o.}\simeq 3$
fm. Our computations were carried out with the SV-bas version of the
Skyrme force~\cite{2009PhRvC..79c4310K}. The magnetic field direction
was chosen along the $z$ direction, but other directions were also
tested to produce identical results. Initially nuclei with $B=0$ were
computed, thereafter the magnetic field was incremented with steps
of about 0.25$ \sqrt{\textrm{MeV} ~\textrm{fm}^{-3}}$ as long as the
convergence was achieved for the number of iterations $N_i\le 4000$.
The direction of the $B$ field was chosen always as the $z$ axis of
Cartesian coordinates, unless stated otherwise.

To access the shape and size of the nuclei we examined the radii
$r_\mathrm{rms}$, the total deformation $\beta$ and the  triaxiality
$\gamma$~(for definitions see, e.~g., \cite{Greiner_Maruhn}). Here
$\gamma=0\degree$ refers to a prolate deformed nucleus,
$\gamma=60\degree$ refers to an oblate deformed nucleus, and angles
between $\gamma=0\degree$ and $\gamma=60\degree$ refer to a
deformation in a state between prolate and oblate. If $\beta=0$ the
nucleus is spherical (undeformed) independent of $\gamma$. For
non-zero $\beta$, the nuclei are deformed.

\section{Results}
\label{sec:results}

In this work we report the studies of symmetrical nuclei
$\isotope[12]{C}$, $\isotope[16]{O}$, and $\isotope[20]{Ne}$ in a strong
magnetic field, for which we evaluated the energy levels of neutrons
and protons, their spin and current densities, as well as deformation
parameters. Below we show selected results from these studies, which
highlight the physics of nuclei in strong $B$ fields. 

\subsection{Analytical estimates}

Before discussing the numerical results we provide approximate
analytical formulas for energy level splitting and $z$ components of
angular momentum and spin in terms of Clebsch-Gordan coefficients. 
We adopt the definition of the Clebsch-Gordan coefficients~\cite{Greiner_Maruhn}:
\begin{eqnarray}
\label{eq:CG}
  \left|JMls\right>&=&\sum_{m_lm_s}\left|lm_lsm_s\right>(lsJ|m_lm_sM)\,,
\end{eqnarray}
where $l$ is the orbital angular momentum, $s$ the spin, and 
$J$ the total angular momentum and $m_l$, $m_s$, and $M$ 
are the $z$ components of $l$, $s$, and $J$, respectively. 
The following conditions need to be fulfilled:
\begin{subequations}
\begin{eqnarray}
  m_l+m_s&=&M\,,\label{sum_m}\\
  |l-s|&\leq&J\leq l+s\,.\label{sum_j}
\end{eqnarray}
\end{subequations}
For nucleons  $s=1/2$ always and, therefore, $m_s=\pm1/2$,
$l$ is a non-negative integer number,  and $m_l$ is an integer number, 
which can be positive, negative, or zero. Thus, $J$ assumes non-negative 
half-integer numbers; $M$ assumes half-integer numbers, 
which can be positive or negative. Writing out formula  \eqref{eq:CG}
for the lowest states of a nucleus and inserting the values of the
Clebsch-Gordan coefficients we find for the $s$ states 
\begin{subequations}
\bea
s_{1/2} (M=-\tfrac12):\quad\left|\tfrac12,-\tfrac12,0\right>&=&\left|0,+0,-\tfrac12\right>,\label{s_m}\\
 s_{1/2} (M=+\tfrac12):\quad\left|\tfrac12,+\tfrac12,0\right>&=&\left|0,+0,\tfrac12\right>,\label{s_p}
\eea
\end{subequations}
where on both side of these equations we omitted the trivial value of
spin $s=1/2$ appearing in Eq.~\eqref{eq:CG}. For the $p$  states we
find 
\begin{subequations}
\begin{eqnarray}
\label{p3_mm}
&&  p_{3/2} (M=-\tfrac32):\\
  &&\qquad\left|\tfrac32,-\tfrac32,1\right>=\left|1,-1,-\tfrac12\right>,
\nonumber\\
&&  \label{p3_m}  p_{3/2} (M=-\tfrac12):\\
&&\qquad \left|\tfrac32,-\tfrac12,1\right>=
      \tfrac1{\sqrt3}\left|1,-1,\tfrac12\right>
    +\sqrt{\tfrac23}\left|1,+0,-\tfrac12\right>,\nonumber\\
&& \label{p3_p} p_{3/2}(M=+\tfrac12):\\
&&\qquad\left|\tfrac32,+\tfrac12,1\right>=\sqrt{\tfrac23}\left|1,+0,\tfrac12\right>
    +\tfrac1{\sqrt3}\left|1,+1,-\tfrac12\right>,\nonumber\\
&&  \label{p3_pp} p_{3/2}(M=+\tfrac32): \\ 
&&\qquad\left|\tfrac32,+\tfrac32,1\right>=\left|1,+1,\tfrac12\right>,\nonumber\\
 && \label{p1_m} p_{1/2}(M=-\tfrac12):\\
&&\qquad\left|\tfrac12,-\tfrac12,1\right>
=-\sqrt{\tfrac23}\left|1,-1,\tfrac12\right>\
    +\tfrac1{\sqrt3}\left|1,+0,-\tfrac12\right>,\nonumber\\
&&\label{p1_p}
  p_{1/2}(M=+\tfrac12):\\
&&\qquad\left|\tfrac12,+\tfrac12,1\right>
=-\tfrac1{\sqrt3}\left|1,+0,\tfrac12\right>
+\sqrt{\tfrac23}\left|1,+1,-\tfrac12\right>.\nonumber
\end{eqnarray}
\end{subequations}
The states can be divided into mixed states that are linear combinations of certain
$\left|lm_lsm_s\right>$ states and pure states which involve only one
set of these quantum numbers. The energy levels of spin-up and -down
particles are split in a magnetic field. This splitting is given by 
\begin{eqnarray}
  \Delta E_q(J,M,l)&=&\left<J,M,l\middle|\hat  h^{(B)}_q\middle|J,M,l\right>,
\end{eqnarray}
which can be expressed in terms of the Clebsch-Gordan coefficients 
\begin{eqnarray}
\label{eq:Eq}
\Delta E_q(J,M,l)&=&-\sum_{m_l,m_s}\left(l,J|m_l,m_s,M\right)^2\tilde B_q\nonumber\\&&\times\left(m_l\delta_{q,p}+g_q m_s\right). \label{delta_energy_1}
\end{eqnarray}
The $z$-components of the angular momentum and spin can be as well
expressed through the Clebsch-Gordan coefficients according to  
\begin{eqnarray}
\label{eq:Sz}
  \left<S_z(J,M,l)\right>
&=&\hbar\sum_{m_l,m_s}m_s\left(l,J|m_l,m_s,M\right)^2,\\
\label{eq:Lz}
\left<L_z(J,M,l)\right>&=&\hbar\sum_{m_l,m_s}m_l\left(l,J|m_l,m_s,M\right)^2.
\end{eqnarray}
Table~\ref{tab:1} lists values of energy splitting, and $z$ components
of angular momentum and spin according to
Eqs.~\eqref{eq:Eq}-\eqref{eq:Lz} for neutron and proton states. These
analytical results can be compared to the results of the \sky code, to
which we now turn.

\subsection{Energy levels}

\begin{figure}[t]
\begin{center}
\includegraphics[width=10.cm]{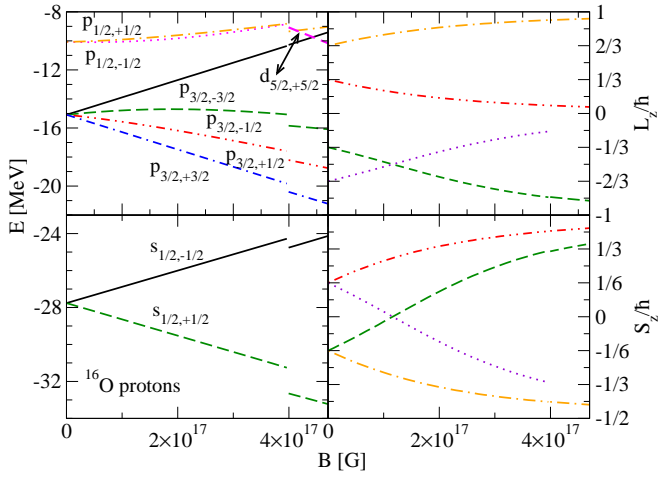}
\caption{ Dependence of proton energy levels of $\isotope[16]{O}$ on
  magnetic field (left panels) and the projections of the angular
  momentum and spin on the $z$ axis. The lower panels show the
  $s$-states, the upper ones the $p$ states and a $d$ state for large
  $B$ fields.  A rearrangement of the energy levels is seen in the
  upper left panel at the magnetic field $B\ge 4\times 10^{17}$
  G. This is related to the rearrangement of the proton levels as
  discussed in the text. }
\label{fig:e_levels_O_protons}
\end{center}
\end{figure}

\begin{table}[bht]
  \caption{Energy splitting and $z$ components of angular momentum and 
    spin for the $s$ and $p$ states. The values of quantities for negative $M$ differ 
    from the corresponding ones for positive $M$ by overall sign and 
    are not shown. }
\begin{tabular}{ccccc}
\hline\hline 
&
$ s_{1/2}(M=\tfrac12 $)&
$ p_{3/2}(M =\tfrac12 $)&
$ p_{3/2}(M =\tfrac32 $)&
$ p_{1/2}(M =\tfrac12 $)\\
\hline 
\\
$\frac{\Delta E_n}{g_nB_n} $ &
$-\tfrac12$&
$-\tfrac16 $& 
$-\tfrac12 $&
$\tfrac16   $
\\
$\frac{\Delta E_p}{g_pB_p}$&
$-\tfrac12  $                              &
$-\tfrac13 (g_p^{-1}+\frac{1}{2})$ & 
$-(g_p^{-1}+\frac{1}{2}) $             &
$-\tfrac23 (g_p^{-1}-\frac{1}{2})$
\\
$\langle L_z\rangle/\hbar $ &
$ 0 $ &
$\tfrac13 $ & 
$1 $             &
$\tfrac23 $\\
$\langle S_z\rangle/\hbar$   &
$ \tfrac12 $ &
$\tfrac16 $ & 
$\tfrac12 $  &
$-\tfrac16 $\\\\
\hline 
\end{tabular}
\label{tab:1}
\end{table}
\begin{figure}[h]
\begin{center}
\includegraphics[width=10cm]{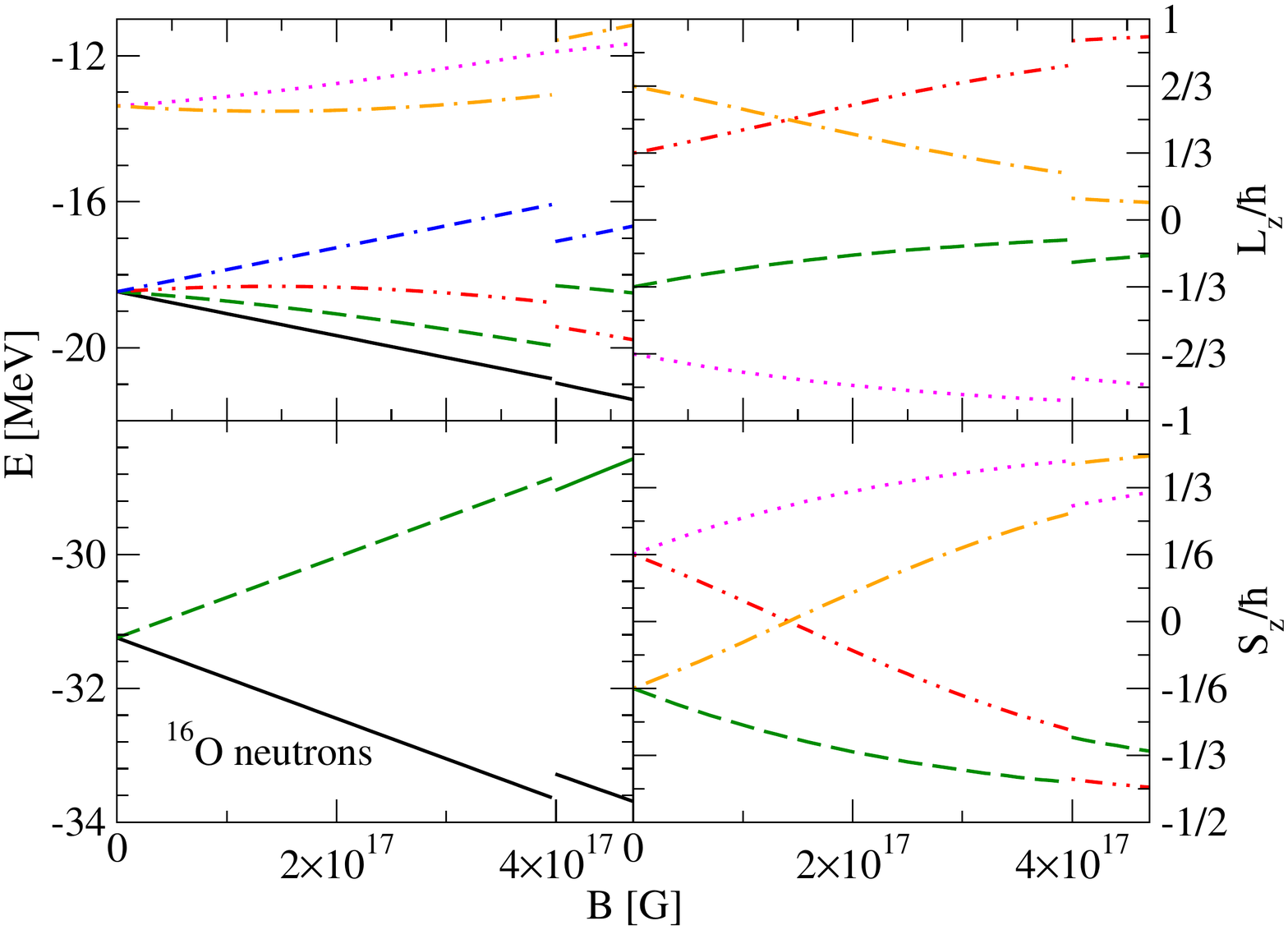}
\caption{Same as  Fig.~\ref{fig:e_levels_O_protons}, but for
  neutrons. }
\label{fig:e_levels_O_neutrons}
\end{center}
\end{figure}

\begin{figure}[t]
\begin{center}
\includegraphics[width=10cm]{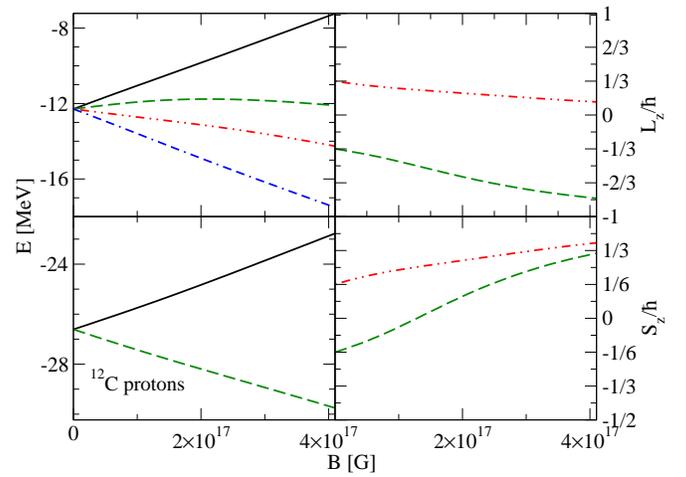}
\caption{Same as  Fig.~\ref{fig:e_levels_O_protons}, but for 
  protons in $\isotope[12]{C}$. No rearrangement of levels is observed
for this nucleus.}
\label{fig:e_levels_C_protons}
\end{center}
\end{figure}

\begin{figure}[h]
\begin{center}
\includegraphics[width=10cm]{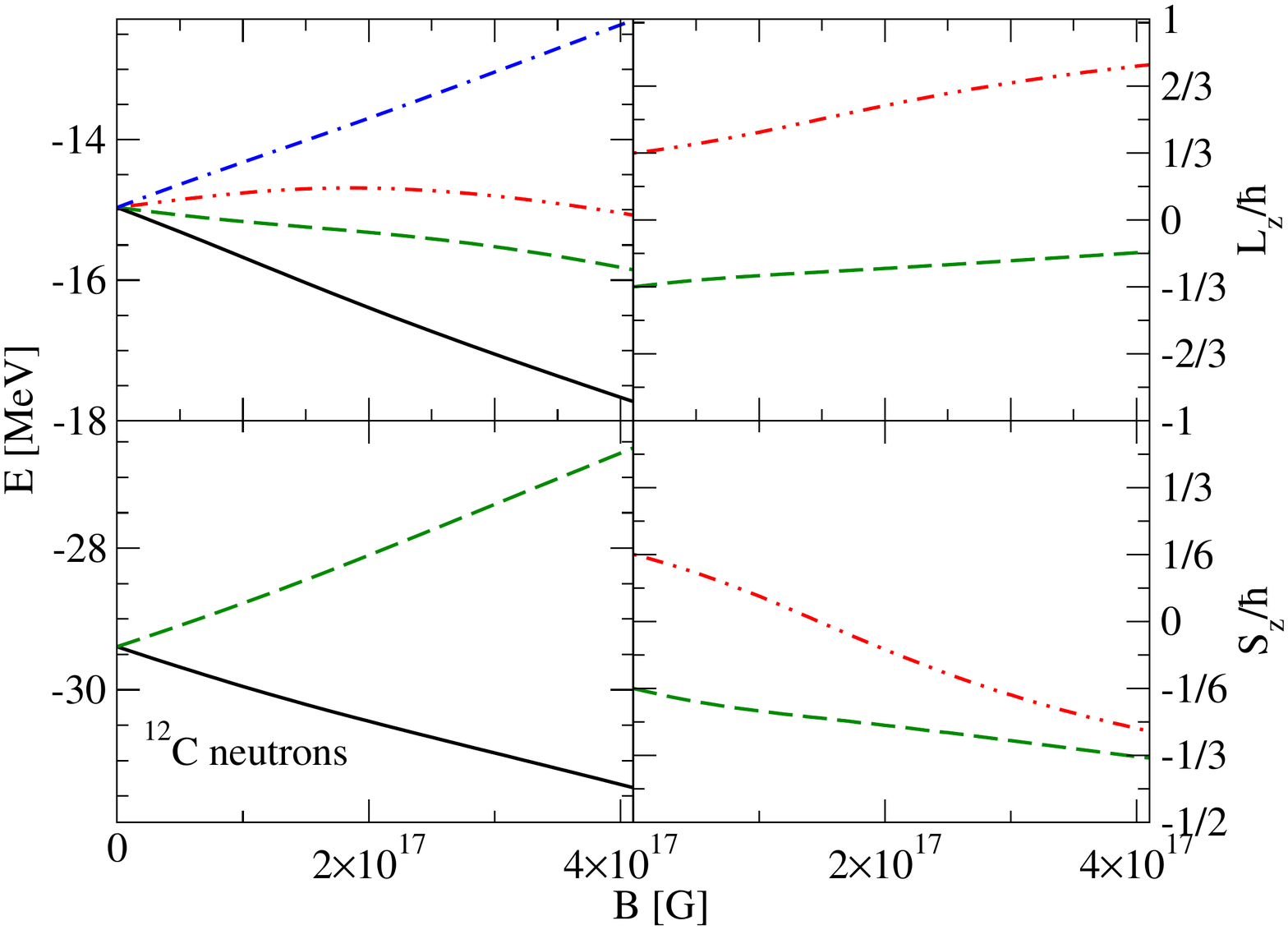}
\caption{Same as Fig.~\ref{fig:e_levels_C_protons}, but for 
  neutrons. }
\label{fig:e_levels_C_neutrons}
\end{center}
\end{figure}
The states of protons and neutrons in $\isotope[16]{O}$ nucleus are
shown in Figs.~\ref{fig:e_levels_O_protons} and
\ref{fig:e_levels_O_neutrons} as a function of the magnetic field. For the
fields $B\ge 10^{16}\,$G the split in the energy levels of spin-up and
spin-down states becomes sizable (of the order of the nuclear scale
MeV). For the $s$ states the splitting increases linearly with the
magnitude of magnetic field.  For $p$ states this dependence is
more complicated. Below the critical field
$B_{c}[\isotope[16]{O}]=4.0\times 10^{17}\,$G, the filling of the
states corresponds to that of the lowest available states of the
harmonic oscillator. However, for  fields larger than
$B_{c}[\isotope[16]{O}]\,$ the occupation pattern for protons
changes: the $d_{5/2},M=+5/2$ state becomes occupied instead of the
$p_{1/2},M=-1/2$ state (which is not shown above this field value) in
Fig.~\ref{fig:e_levels_O_protons}. We also find that above the
critical field $B_{c}[\isotope[16]{O}]$ the energy levels undergo an
abrupt rearrangement.
For the pure states (all $s$ states and $p$ states with $M=\pm3/2)$,
we obtain a good agreement between the numerical and analytical
results for the energy levels. For the mixed states the discrepancy 
between the numerical and analytical results is below 10$\%$ of
the energy of the corresponding level. 
The computations of the energy levels for the $\isotope[12]{C}$ nucleus
show the same basic features seen already in the case of
$\isotope[16]{O}$, see Figs.~\ref{fig:e_levels_C_protons} and \ref{fig:e_levels_C_neutrons}. 
In this case the numerical and analytical differ
by at most  10$\%$ of the energy of the levels, the discrepancy
increasing with the field. The same applies to the case of the
$\isotope[20]{Ne}$ nucleus, but classification of the levels in this
case is more complex because apart from the $1s$ and $1p$ states two
$1d_{5/2}$ states should be filled. This nucleus is deformed in the
ground state and the axis of the deformation may not coincide with the
direction of the magnetic field, in which case there are no states
with half-integer values of $(\langle L_z\rangle +\langle
S_z\rangle)/\hbar$.
In addition to the energy levels, we have also computed the $z$
components of orbital angular momentum $\left<L_z\right>$ and spin
$\left<S_z\right>$ of neutrons and protons as functions of the
magnetic field.  The results are shown in
Figs.~\ref{fig:e_levels_O_protons} and~\ref{fig:e_levels_O_neutrons}.
For the $s$ states, defined in Eqs.~\eqref{s_m} and \eqref{s_p}, as
well as pure $p$ states \eqref{p3_mm} and \eqref{p3_pp} we obtain
integer or half integer numbers for $\left<L_z\right>/\hbar$ or
$\left<S_z\right>/\hbar$ which are identical to the quantum numbers
$m_l$ or $m_s$, respectively, are independent of the magnetic field,
and therefore are not shown in Figs.~\ref{fig:e_levels_O_protons}
and~\ref{fig:e_levels_O_neutrons}.

For mixed states given by Eqs.~\eqref{p3_m}, \eqref{p3_p},
\eqref{p1_m} and \eqref{p1_p}, $\left<L_z\right>$ and
$\left<S_z\right>$ change as functions of the magnetic field as seen
from these figures. For these states at any magnetic field
$M=m_l+m_s=(\left<L_z\right>+\left<S_z\right>)/\hbar$ is a good
quantum number.  In the limit of weak magnetic fields the angular momentum
and spin are coupled via the {\it l-s} coupling.  Then
$\left<L_z\right>$ and $\left<S_z\right>$ for each state are given according to
Table~\ref{tab:1}. Because of the {\it l-s} coupling the vectors of
$\vecL$ and $\vecS$ are not aligned with the magnetic field
separately.  The influence of the weak magnetic field on the system is
described by the Zeeman effect.  In the regime of strong magnetic
fields the {\it l-s} coupling is ineffective, i.e., the orbital
angular momentum $l$ and the spin $s$ couple separately to the
magnetic field. In this case, the mixed states  reach asymptotically the
following non mixed states:
\begin{subequations}
\label{clebsch_gordan_coefficients_high_b}
\begin{eqnarray}
  p_{3/2} (M=-\tfrac12):&\quad&\left|1,+0,-\tfrac12\right>\,,\label{p3_m_high_b}\\
  p_{3/2}(M=+\tfrac12):&\quad&\left|1,+1,-\tfrac12\right>\,,\label{p3_p_high_b}\\
  p_{1/2}(M=-\tfrac12):&\quad&\left|1,-1,\tfrac12\right>\,,\label{p1_m_high_b}\\
  p_{1/2}(M=+\tfrac12):&\quad&\left|1,+0,\tfrac12\right>\,.\label{p1_p_high_b}
\end{eqnarray}
\end{subequations}
Thus we observe a smooth transition from the {\it l-s} to the separate
{\it l-B} and {\it s-B} coupling as the magnetic field is increased,
which can be viewed as a transition from the Zeeman to the Paschen-Back
effect.  Finally we note that for the $d_{5/2} (M=+5/2)$ state
$\left<L_z\right>/\hbar=m_l=2$ and $\left<S_z\right>/\hbar=m_s=1/2$
assume constant values for all magnetic fields and are therefore not
shown in Fig.~\ref{fig:e_levels_O_protons}. The key features observed
for components $\left<L_z\right>$ and $\left<S_z\right>$ of
$\isotope[16]{O}$ persist in the case of $\isotope[12]{C}$
and we do not show them here.

\subsection{Spin and current densities} 

\begin{figure}[tbh]
\hskip 2cm 
\includegraphics[width=5cm,height=5cm,angle=0]{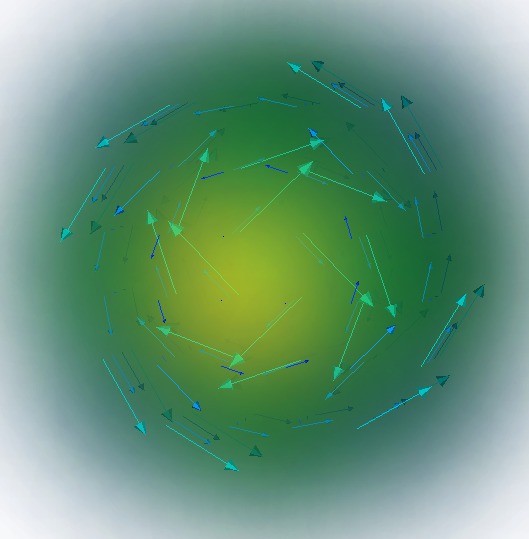}
\includegraphics[width=2cm,height=3.8cm,angle=0]{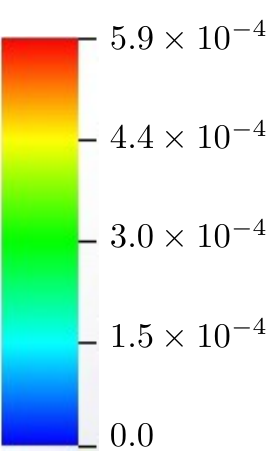}
\includegraphics[width=5cm,height=5cm,angle=0]{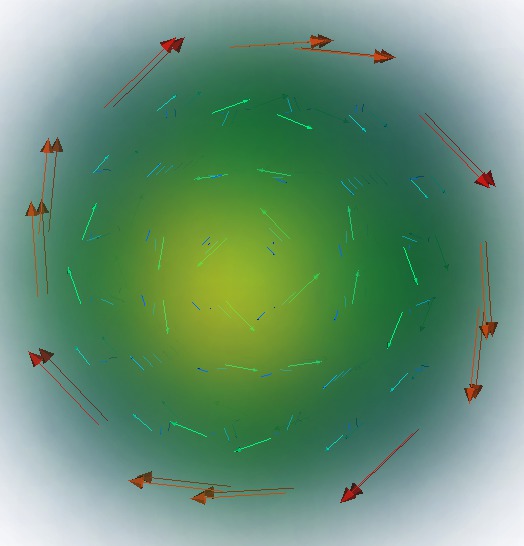}
\includegraphics[width=2cm,height=3.8cm,angle=0]{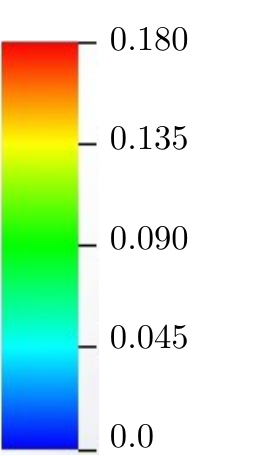}
\includegraphics[width=5cm,height=5cm,angle=0]{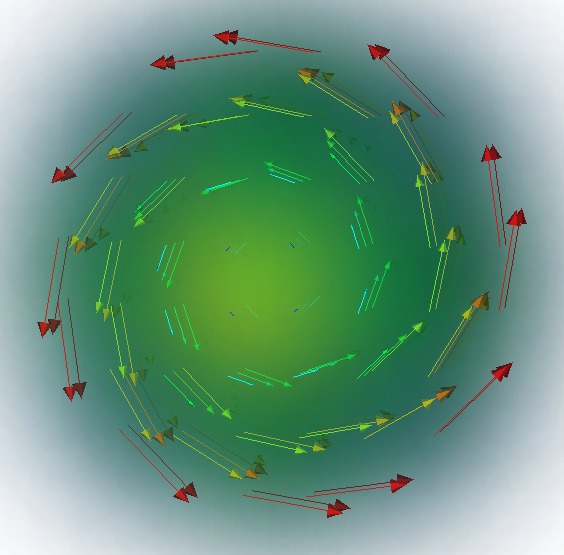}
\caption[] {Velocity distribution for neutrons (upper panel) and
  protons (middle and bottom panels) in $\isotope[16]{O}$. The upper
  and middle panels have the same color coding show on the left and
  correspond to magnetic field $B=3.9\times 10^{17}\,$G. The bottom
  panel with its color coding corresponds to magnetic field
  $B=4.1\times 10^{17}\,$G $ > B_{c}[\isotope[16]{O}]$.The background
  shows the density distribution with maximum 0.155 fm$^{-3}$ at the
  center (yellow color) to the boundary where the density drops to
  zero (dark blue). The velocity is shown in units of speed of light.
}
\label{fig:current_O}
\end{figure}
\begin{figure}[tbh]
\includegraphics[width=2cm,height=3.8cm,angle=0]{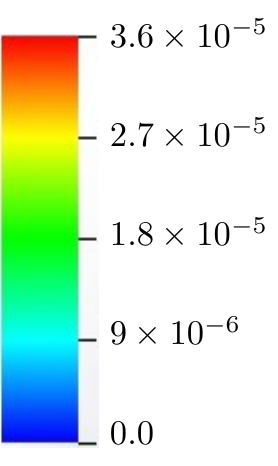}
\includegraphics[width=5cm,height=5cm,angle=0]{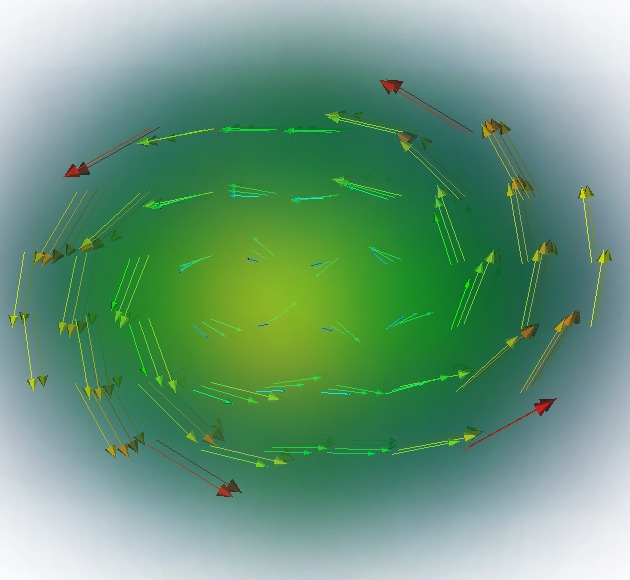}
\includegraphics[width=2cm,height=3.8cm,angle=0]{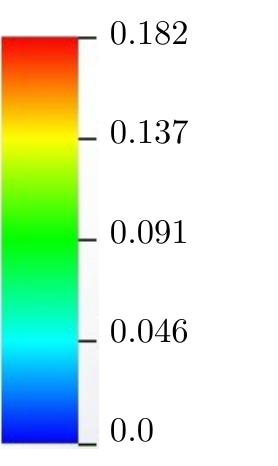}
\includegraphics[width=5cm,height=5cm,angle=0]{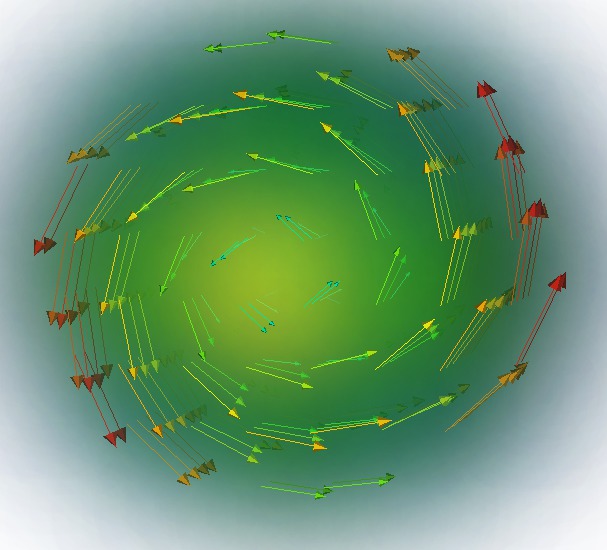}
\caption[] {Same as  Fig.~\ref{fig:current_O}, but for protons in \isotope[20]{Ne}
  and for fields $B=4.0\times 10^{13}\,$G (top panel) and $B=4.9 
  \times 10^{17}\,$G (bottom panel). 
}
\label{fig:current_N}
\end{figure}

We have extracted the current and spin densities as a functions of the
magnetic field for the $\isotope[12]{C}$, $\isotope[16]{O}$, and
$\isotope[20]{N}$ nuclei. It is more convenient to show the collective
flow velocity instead of the currents by dividing these with the
particle density. The velocity distribution in $\isotope[16]{O}$ is
shown in Fig.~\ref{fig:current_O}. The top and middle panels compare
the velocity distribution for neutrons and protons for the field
$B=3.9\times 10^{17}\,$G; the background shows the density
distribution within the nucleus. It is seen that the magnitude of
the proton current is by a factor of 4 larger than the neutron current. We
also observe that the proton and neutron currents are counter-moving
and concentrated around the surface of the nucleus.  Note that the
current density is shown in the plane orthogonal to the field.
In the case of $\isotope[12]{C}$ we find current patterns of
neutrons and protons similar to those seen in $\isotope[16]{O}$. 
The currents are counter-moving for neutrons and protons,
their magnitudes increase with the field value and are mostly
concentrated at the surface of the nuclei. Because there is no
rearrangement seen for $\isotope[12]{C}$ this increase, in contrast to
$\isotope[16]{O}$, is gradual. 
The velocities of protons in \isotope[20]{Ne} are shown in
Fig.~\ref{fig:current_N} for two magnetic fields
$B=4.0\times 10^{13}\,$G and the maximal studied field
$B=4.9 \times 10^{17}\,$G. The four orders of magnitude increase in the
field leads to an increase in the maximal current by about the same
factor. The current is concentrated at the
surface of the field; note that the magnetic field induces a change in
the shape of the nucleus (density distribution) which in turn affects
the pattern of currents, which are more circular for a larger magnetic
field. 

\begin{figure}[t]
\hspace{2cm}
\includegraphics[width=5cm,height=5cm,angle=0]{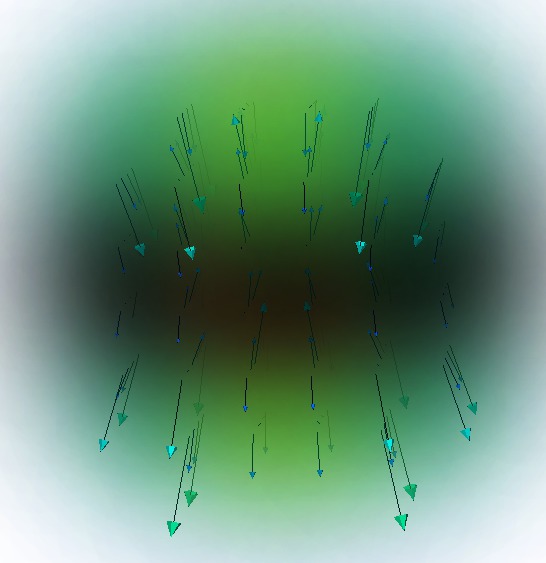}\\
\includegraphics[width=2cm,height=3.8cm,angle=0]{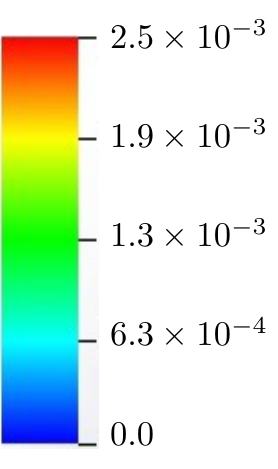}
\includegraphics[width=5cm,height=5cm,angle=0]{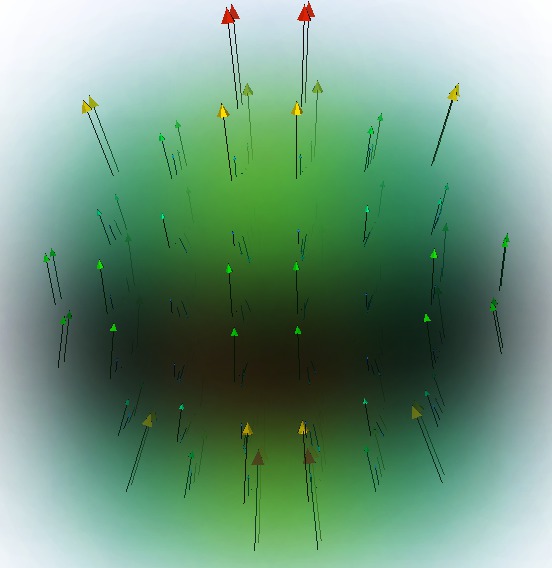}\\
\includegraphics[width=2cm,height=3.8cm,angle=0]{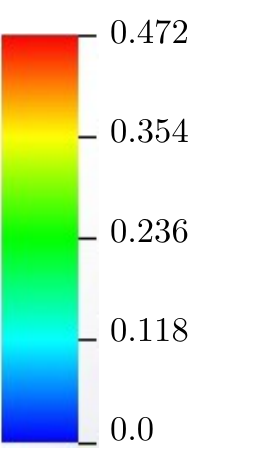}
\includegraphics[width=5cm,height=5cm,angle=0]{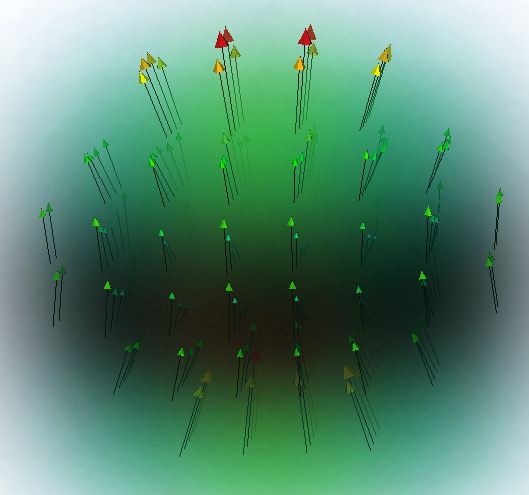}
\caption[] {The ratio of the spin density to the particle density 
 for neutrons (upper panel) and
  protons (middle and bottom panel) for \isotope[16]{O} nucleus. The
  magnetic field $z$ axis is directed from bottom to top.  The upper
  and middle panel have the same color coding shown on the left and
  correspond to magnetic field $B=3.9\times 10^{17}\,$G. The bottom
  panel with its color coding corresponds to magnetic field
  $B=4.1\times 10^{17}\,$G.  }
\label{fig:spin_O}
\end{figure}
\begin{figure}[t]
\includegraphics[width=2cm,height=3.8cm,angle=0]{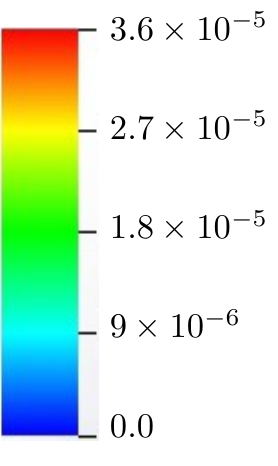}
\includegraphics[width=5cm,height=5cm,angle=0]{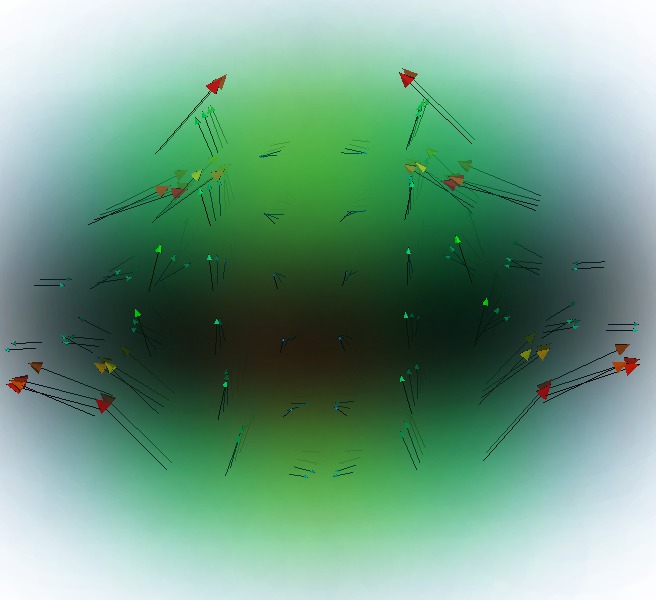}
\\
\includegraphics[width=2cm,height=3.8cm,angle=0]{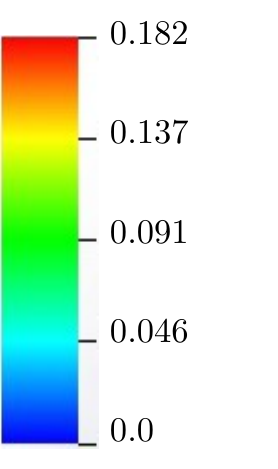}
\includegraphics[width=5cm,height=5cm,angle=0]{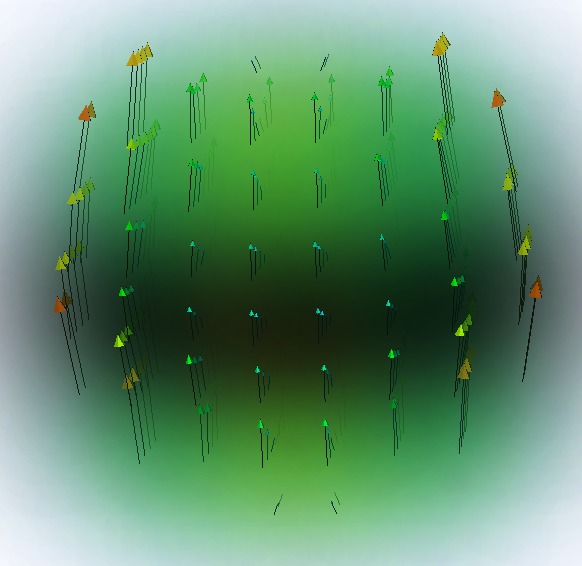}
\caption[] {Same as in Fig.~\ref{fig:spin_O} for protons in \isotope[20]{Ne}
  and for fields $B=4.0\times 10^{13}\,$G (top panel) and $B=4.9
  \times 10^{17}\,$G (bottom panel).}
\label{fig:spin_N}
\end{figure}
We consider now the spin density of neutrons and protons in the nuclei
under consideration and the effect of the spin interaction with the
magnetic field. The spin-density for \isotope[16]{O} is shown in
Fig.~\ref{fig:spin_O}. The spins of neutrons and protons are
anti-parallel as expected from Eq.~\eqref{sphmod}. In the case of
\isotope[16]{O} the spin polarization increases abruptly for protons
because of the rearrangement of the levels. For weak magnetic fields
the {\it l-s} coupling is dominant therefore the alignment of the
spins is not pronounced; for strong magnetic fields the spins are
aligned with the $z$-axis. We find that in the case of \isotope[12]{C}
the spin alignment is more pronounced than for \isotope[16]{O}, i.e.,
\isotope[12]{C} is more polarizable.  In Fig.~\ref{fig:spin_N} we show
the spin polarization in the \isotope[20]{Ne} nucleus, where the new
feature is the deformation of the nucleus in the ground state in the
absence of a magnetic field. For small magnetic fields a clear
evidence of two symmetry axes is present. For larger magnetic fields
the spin polarization by the magnetic field directs the spin alongs
the field axis ($z$  direction). We note that in all cases protons and
neutron show quantitatively similar polarization (the minor difference
coming from different values of their $g$ factors).

\subsection{Nuclear shapes}

Finally, we want to consider the shape and the size of the
\isotope[16]{O} nucleus for non-zero magnetic fields. For
$B<B_c[\isotope[16]{O}]\,$, the lowest states of the harmonic
oscillator are filled and the shape of the nucleus is spherical. Its
radius is $r_\mathrm{rms}=2.69\,$fm. For $B>B_c[\isotope[16]{O}]\,$
the nucleus is deformed with deformation parameters $\beta=0.1$ and
$\gamma=60\degree$, which implies that the deformation is oblate. The
mean radius is $r_\mathrm{rms}=2.72\,$fm in this case. The
redistribution of neutron and proton states above the critical field
has the effect of slightly deforming the nucleus from its spherical
shape. We stress that the redistribution is found to be abrupt and
therefore the change in the shape of the nucleus is abrupt as well.

In the case of the \isotope[12]{C} nucleus we find that the shape of
the nucleus does not change much with increasing magnetic field.  For
a zero field it is spherical symmetric with $r_\mathrm{rms}=2.47\,$fm,
increasing slightly to 2.51\,fm for
$B_c=4.1\times 10^{17}\,$G. Increasing the magnetic field also results
in a smooth deformation from $\beta=0$ at $B=0$ to $\beta=0.071$ for
$B_c$. The deformation is always oblate with $\gamma=60\degree$.

The deformation of the \isotope[20]{Ne} nucleus with the magnetic
field is continuous and in contrast to the other examples above, the
nucleus is deformed for a vanishing magnetic field.  We find that its
radius $r_\mathrm{rms}$ slightly decreases from 2.93\,fm to 2.87\,fm
with the magnetic field.  However, the $\beta$ parameter decreases from
0.32 to 0.15, i.e., to a value which is less than half of the original
one. This is the only example of a continuous significant change in 
deformation as a function of magnetic field. 
The parameter $\gamma$
starts at $0\degree$ for $B=0$, but increases asymptotically to
$11\degree$ implying a change from a purely prolate deformed nucleus
to a mainly prolate deformed one. This evolution of \isotope[20]{Ne}
nucleus from deformed to the  more spherical shape is visualized 
in Fig.~\ref{fig:current_N}.

\section{Conclusions}
\label{sec:conclusions}

We have performed numerical computations of $\isotope[12]{C}$,
$\isotope[16]{O}$, and $\isotope[20]{Ne}$ nuclei in strong magnetic
fields using an extension of the \sky code, which solves Hartree-Fock
equations on a three-dimensional grid in a strong magnetic field. The code is based on
the Skyrme density functional. Common features found for all three
nuclei are (i) the splitting of energy states, which are on the order of
MeV for fields $B\sim 10^{17}\,$G; (ii) the increase in
spin polarization along the magnetic field as the field is increased,
which is characterized by a transition from a regime where $l$-$s$
coupling is dominant to a regime where $l$ and $s$ couple directly to
the magnetic field; and (iii) an increase in the flow-velocity in the
plane orthogonal to the field with increasing magnetic fields.  A
number of features are peculiar to specific nuclei and are listed
below:
\begin{itemize}
\item A rearrangement of energy levels in $\isotope[16]{O}$ nucleus is
  observed at a critical field $4.1\times 10^{17}\,$G, which is
  accompanied by an abrupt increase in the magnetization of the
  nucleus and an increase in the velocity flow. This
  also leads to deformation of the nucleus from
  its original spherical shape at vanishing value of the field.
\item The $\isotope[12]{C}$ nucleus does not change its shape in the
  magnetic field and there are no energy level rearrangements as seen in
  $\isotope[16]{O}$. It is found to be more easily polarizable
  than the heavier nuclei.
\item The $\isotope[20]{Ne}$ nucleus is deformed in the ground
  state. It undergoes significant continuous change in its shape as
  the magnetic field is increased. The deformation is diminished by a
  factor of 2 for fields  $B\simeq 4.1\times 10^{17}\,$G. 
\item We have shown that a simple analytical model which fills in the
  harmonic oscillator states in the magnetic field accounts well for the
  energy splitting of $\isotope[12]{C}$ and $\isotope[16]{O}$ nuclei
  as well as their angular momenta and spin projections. In the case
  of $\isotope[20]{Ne}$ the analytical model is less reliable; it can
  reproduce qualitatively features obtained with the \sky code;
  however, because of the nuclear deformation, the magnetic field needs
  to be directed along the $x$-axis instead of the $z$-axis.
\end{itemize}

Phenomenologically the most important aspect of these findings is the
splitting of the levels in nuclei as a function of the $B$ field. When
this splitting is on the order of the temperature it will have an important
impact on all transport processes and on neutrino and photon emission
and absorption, as well as on the reaction rates. 

Looking ahead, we would like to extend these studies to nuclei with
larger mass numbers and beyond the stability valley in the direction
of neutron-rich nuclei that occur in nonaccreting neutron
stars~\cite{PhysRevLett.110.041101}.  The possibility of non-spherical
and extended nuclei (pasta phases~\cite{2016PhRvC..93e5801S}) can also
be considered in this context.


\acknowledgments

M. S. acknowledges support from the HGS-HIRe graduate program at
Frankfurt University. A. S. is supported by the Deutsche
Forschungsgemeinschaft (Grant No. SE 1836/3-1) and by the NewCompStar
COST Action MP1304.

\end{document}